\documentclass[a4, 10pt,twoside,twocolumn]{elsart}							%TWO_COLUMN

\journal{Nucl. Instrum. Meth. A}											%TWO_COLUMN
\date{}																	    %TWO_COLUMN
\usepackage{graphicx}
\usepackage{amssymb}
\usepackage{amsmath}

\newcommand{\spaceafterfigurecaption}{\vspace{9mm}}
\newcommand{\extraspaceafterfig}{\vspace{0mm}}

\newcommand{\lowsp}{\vspace{-1.5mm}}
\newcommand{\fns}{\footnotesize}

\newcommand{\ftMicos}{\footnote{Micos Gmbh, Freiburger Stra\ss{}e 30, 79427
Eschbach (Germany). %http://www.micos.ws
}}
\newcommand{\ftMicron}{\footnote{Micron Technology Inc., Boise, Idaho (USA).
The sensor was produced before by Photobit Inc, Pasadena, CA (USA).
}}
\newcommand{\ftMikrotron}{\footnote{Mikrotron GmbH, Landshuter Str.20-22
D-85716 Unterschleissheim (Germany).
}}
\newcommand{\ftNational}{\footnote{National Instruments Corporation,
11500 N MoPac Expwy, Austin, TX (USA). %http://www.ni.com
}}
\newcommand{\ftMatrox}{\footnote{Matrox Electronic Systems Ltd., 1055 St.
Regis Blvd., Dorval, Quebec (Canada). %http://www.matrox.com/imaging/
}}
\newcommand{\ftOriental}{\footnote{Oriental Motor Co. Ltd., 6-16-17 Ueno,
Taito-ku, Tokyo (Japan).
}}
\newcommand{\ThickBase}{205\ }
\newcommand{\ThickEmu}{44\ }

\begin{document}

\begin{frontmatter}


%\fbox{\textbf{Author list taken from NIM-A551(2005)261: to be updated !!!}}

\title{Hardware performance of a scanning system for high speed
analysis of nuclear emulsions}
\author[LY]{L. Arrabito}, 
\author[SA]{E. Barbuto}, 
\author[SA]{C. Bozza}, 
\author[NA]{S. Buontempo}, 
\author[BO]{L. Consiglio}, 
\author[NA]{D. Coppola}, 
\author[BO]{M. Cozzi}, 
\author[BE]{J. Damet\thanksref{LAP}},
\thanks[LAP]{Now at LAPP, IN2P3-CNRS and Universit\`e de Savoie, Annecy, France.}
\author[NA]{N. D'Ambrosio}, 
\author[NA]{G. De Lellis}, 
\author[BA]{M. De Serio}, 
\author[NA]{F. Di Capua}, 
\author[BO]{D. Di Ferdinando},
\author[AQ,GS]{N. Di Marco}, 
\author[GS]{L.S. Esposito}, 
\author[BO]{G. Giacomelli}, 
\author[SA]{G. Grella}, 
\author[NE]{M. Hauger},
\author[NE]{F. Juget},
\author[BE]{I. Kreslo},
\author[BO]{M. Giorgini}, 
\author[BA]{M. Ieva}, 
\author[LY]{I. Laktineh}, 
\author[LY]{K. Manai},
\author[BO]{G. Mandrioli}, 
\author[NA]{A. Marotta}, 
\author[BO]{S. Manzoor},
\author[NA]{P. Migliozzi}, 
\author[AQ,GS]{P. Monacelli},
\author[BA]{M.T. Muciaccia},
\author[BA]{A. Pastore}, 
\author[BO]{L. Patrizii}, 
\author[NA]{C. Pistillo}, 
\author[BO]{M. Pozzato},
\author[LY]{P. Royole-Degieux}, 
\author[SA]{G. Romano}, 
\author[RM]{G. Rosa},
\author[BE]{N. Savvinov},
\author[RM]{A. Schembri}, 
\author[NA]{L. Scotto Lavina},
\author[BA]{S. Simone},
\author[BO]{M. Sioli}, 
\author[SA]{C. Sirignano},
\author[BO]{G. Sirri\corauthref{COR1}},
\corauth[COR1]{Corresponding author: Tel.: +39-051-2095228; fax: +39-051-2095269.} 
\ead{gabriele.sirri@bo.infn.it}
\author[NA]{G. Sorrentino}, 
\author[NA]{P. Strolin}, 
\author[NA]{V. Tioukov},
\author[BE]{T. Waelchli}
\address[LY]{\fns IPNL, IN2P3-CNRS, Universit\'e C. Bernard Lyon-1, 69622 Villeurbanne, France}
%\address[LY]{\fns IPNL, IN2P3-CNRS, Universit\`e C. Bernard Lyon - I, 69622 Villeurbanne, France}
%\address[LY]{\fns Institut de Physique Nucléaire de Lyon, IN2P3-CNRS, Université Claude Bernard, F-69622 Villeurbanne, France}
\lowsp\address[SA]{\fns Dip. di Fisica dell'Universit\`a di Salerno and INFN, 84081 Baronissi (SA), Italy}
\lowsp\address[NA]{\fns Dip. di Fisica dell'Universit\`a ``Federico II'' and INFN, 80126 Napoli, Italy}
\lowsp\address[BO]{\fns Dip. di Fisica dell'Universit\`a di Bologna and INFN, 40127 Bologna, Italy}
\lowsp\address[BE]{\fns Laboratory for High Energy Physics, University of Bern, 3012 Bern, Switzerland}
\lowsp\address[BA]{\fns Dip. di Fisica dell'Universit\`a di Bari and INFN, 70126  Bari, Italy}
\lowsp\address[AQ]{\fns Dip. di Fisica dell'Universit\`a dell'Aquila and INFN, 67100 L'Aquila, Italy}
\lowsp\address[GS]{\fns Laboratori Nazionali del Gran Sasso dell'INFN, 67010 Assergi (L'Aquila), Italy}
\lowsp\address[NE]{\fns Institut de Physique, Universit\'e de Neuch\^atel, 2000 Neuch\^atel, Switzerland}
%\lowsp\address[NE]{\fns Neuch\^atel University, Neuch\^atel, Switzerland}
\lowsp\address[RM]{\fns Dip. di Fisica dell'Universit\`a ``La Sapienza'' and INFN, 00185 Roma, Italy}

\vspace{5mm}\textsl{ \normalsize This paper is dedicated to the memory of our colleague N. Armenise}
\vspace{-3mm}
%______________________________________________________________________________
%

\begin{abstract}
The use of nuclear emulsions in very large physics experiments is now possible thanks to 
the recent improvements in the industrial production of emulsions and to the development of fast automated microscopes. In this paper the hardware performances of the \textit{European Scanning System} (ESS) are described. The ESS is a very fast automatic system developed for the mass scanning of the emulsions of the OPERA experiment, which requires microscopes with scanning speeds of $\sim20$ cm$^2$/h in an emulsion volume of \ThickEmu $\mu$m thickness.
\end{abstract}

\end{frontmatter}

%______________________________________________________________________________
%

\section{Introduction\label{Introduction}}

Nuclear emulsions record tracks of charged particles with an accuracy of less 
than 1~$\mu$m. Thanks to this feature, they were largely used in nuclear and particle physics experiments during the last century\,\cite{powell,barkas} and
they are still successfully used nowadays, especially in experiments involving 
short-lived particles\,\cite{Ushida:1984vd,chorus,DONUT}.
   
The amount of emulsions used in the early experiments was relatively small and 
the measurements were made manually. Significant improvements in the emulsion 
technique and the development of fast automated scanning systems during the last two decades\,\cite{TS,SUTS,SYSAL} have made possible the use of nuclear emulsions in large scale experiments.
   
OPERA is a long baseline experiment\,\cite{OPERA1} (located in 
the INFN Gran Sasso Underground Laboratories)  which will use a
hybrid detector to search for $\nu_{\mu} \rightarrow\nu_{\tau}$ oscillations in the parameter range suggested by atmospheric neutrino 
experiments\,\cite{Fukuda:1998mi,Ambrosio:1998wu,Allison:2005dt}. The goal is to observe  the appearance of $\tau$ leptons in a pure $\nu_\mu$ beam produced at the CERN-SPS (the CNGS neutrino beam\,\cite{CNGS}). The $\tau$ leptons are 
identified through the direct detection of their decays that, at the CNGS energies, are at distances of $\sim1$\,mm from the production point. 
Therefore, a high precision detector is needed. On the other hand, given the 
smallness of the neutrino oscillation probability and of the neutrino 
cross-section a very massive detector ($\mathcal{O}(kton)$) is mandatory. The 
Emulsion Cloud Chamber (ECC)\,\cite{ECC}, a sandwich of dense passive material 
(Pb) sheets, acting as target,  interleaved with emulsion sheets, acting as high precision trackers, satisfies the need of both a large mass and a high precision tracking capability. The ECC technique has lead to the first observation of $\nu_{\tau}$ interactions by the DONUT experiment\,\cite{DONUT}.
   
The OPERA detector is a hybrid system consisting of electronic trackers, muon spectrometers and a massive lead-emulsion target segmented into ECC bricks, each  with size $12.7{\times}10.2{\times}7.5$~cm$^3$ and
consisting of 57 emulsion sheets interleaved with 1 mm lead plates. Emulsion sheets are made of two $\ThickEmu\mu$m thick films (including $1$\,$\mu$m insensitive protective layer) coated on both sides of a $\ThickBase\mu$m thick plastic support \cite{OpFilm}.
   
Electronic detectors are used to identify the brick where the neutrino 
interaction occurred. The brick is then removed and two external emulsion sheets placed downstream of the brick (\textit{changeable sheets}) are 
promptly detached and analyzed to confirm the interaction. 

With the CNGS neutrino beam\,\cite{CNGS} at its nominal intensity, $\sim 30$ neutrino selected interactions per day are expected. Therefore, $\sim 2000$ emulsion sheets per day must be (partially) scanned in order to find the vertex and analyze the event. In total, $\sim6000$\,$\mbox{cm}^2$ per day ($\sim200$\,$\mbox{cm}^2$ per brick) have to be analyzed with a sub-micrometric precision per 5 
years of data taking ($\gtrsim 30000$ neutrino interactions). 
   
The need for a daily scanning of all neutrino interactions comes from the goal to analyze in ``real'' time the events and, for some decay topologies, remove other ECC bricks for a more refined kinematical analysis. Consequently, a very fast automatic scanning system is needed to cope with the daily analysis of the large number of emulsion sheets associated with neutrino interactions. In order to have a reasonable number of microscopes ($\sim$ 1 microscope/brick/day), the minimum required scanning speed is about $20\,\mathrm{cm^2/h}$ per emulsion layer ($\ThickEmu$ $\mu$m thick). It corresponds to an increase in speed by more than one order of magnitude with respect to past systems\,\cite{TS,SYSAL}.
   
For this purpose new automatic fast microscopes have been developed: the \textit{European Scanning System} (ESS) \cite{ESSsw,ESS2} and the S-UTS in Japan \cite{SUTS}. 
   
In this paper the features and performances of the ESS hardware are discussed. High speed particle tracking for the ESS is described in Ref.~\cite{ESSsw,ESS2}, precision measurements in Ref.~\cite{PreMea}, alignments with cosmic ray muons in Ref.~\cite{Barbuto:2004da}, items related to event analysis in Ref.~\cite{Fedra}.

%______________________________________________________________________________
%

\section{The design of the European Scanning System}
\begin{figure*}[tbp]
 \begin{center}
  \resizebox{0.85\textwidth}{!} {
   \includegraphics[clip]{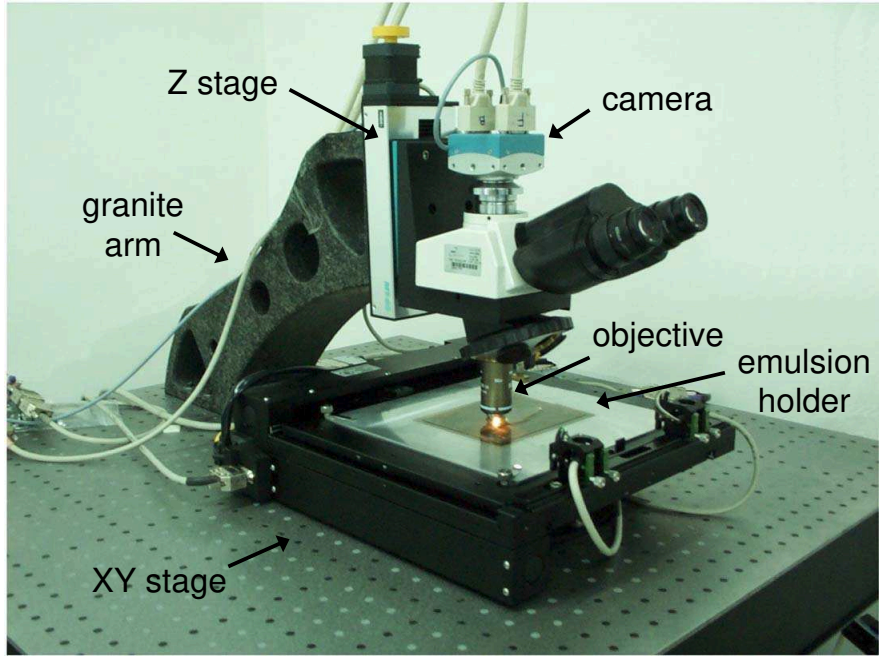} }
  \caption{A photograph of one of the microscopes of the European
  Scanning System.\label{fi:ESS}}
 \end{center} \spaceafterfigurecaption
\end{figure*}

The main components of the microscope shown in Fig.~\ref{fi:ESS} are: (i) a high quality, rigid and vibration-free support table holding the components in a fixed position; (ii) a motor driven scanning stage for horizontal (XY) motion; (iii) a granite arm which acts as an optical stand; (iv) a motor driven stage mounted vertically (Z) on the granite arm for focusing; (v) optics; (vi) digital camera for image grabbing mounted on the vertical stage and connected with a vision processor; (vii) an illumination system located below the scanning table. The emulsion sheet is placed on a glass plate (emulsion holder) and its flatness is guaranteed by a vacuum system which holds the emulsion at a fixed position during the scanning. 

\begin{figure*}[ht]
 \begin{center}
  \resizebox{1.\textwidth}{!} {
     \includegraphics[clip]{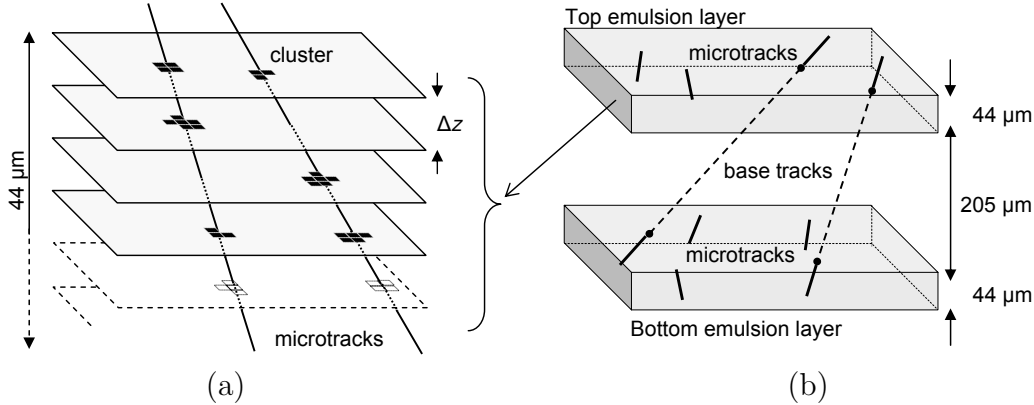} }
    (a) \hspace{0.5\textwidth} (b)
  \caption{(a) Microtrack reconstruction in one emulsion layer by combining clusters belonging to images at different levels. \label{fi:Microtracking} (b) Microtrack connections across the plastic base to form base tracks.\label{fi:Linking} }
 \end{center} \spaceafterfigurecaption
\end{figure*}

By adjusting the focal plane of the objective, the whole \ThickEmu$\mu$m 
emulsion thickness is spanned and a sequence of 15 tomographic images of each field of view, taken at equally spaced depth levels, is obtained. Emulsion images are digitized, converted into a grey scale of 256 levels, sent to a vision processor board, hosted in the control workstation, and analyzed to recognize sequences of aligned grains (clusters of dark pixels of given shape and size). Some of these spots are track grains; others, in fact the majority, are spurious grains 
(\textit{fog}), created in the emulsions by thermal effects and not 
associated to particle tracks. The three-dimensional structure of a track 
in an emulsion layer (\textit{microtrack}) is reconstructed by combining clusters belonging to images at different levels and searching for geometrical alignments (Fig. \ref{fi:Microtracking}a). Each microtrack pair is connected across the plastic base to form the \textit{base track} (Fig. \ref{fi:Linking}b).

A linear fit to these clusters allows the determination of the track position 
and slope\,\cite{ESSsw}.
   
The ESS microscope has been designed according to the following specifications:
\begin{itemize}
\item high-speed computer-controlled precision mechanics for both horizontal and vertical stages with sub-micron accuracy able to move from one field of view to the next in less than 0.1\,s;
\item optical system from standard microscopes, customized to observe the OPERA emulsion sheets which have two emulsion layers on both sides of a plastic support for a total thickness of $\sim 300$ $\mu$m.\footnote{The conditions are completely different from typical ordinary cases: for instance the microscopes normally used in biology to observe few microns thick specimens do not measure the vertical dimension (\textit{image analyzers}).}
\item high-resolution camera interfaced with a high-speed frame grabber and a vision processor able to grab and process images at rates $>$ 350 frames per second (fps).
\end{itemize}

The ESS is based on the use of commercial hardware components or developed in collaboration with specialized companies. The software used for data taking and track reconstruction has a modular structure, providing the flexibility needed to upgrade the system following the technological progress.
   
%______________________________________________________________________________
%
\section{Mechanics}

\subsection{Horizontal stages}
The scanning table and the vertical stage have been developed in collaboration with the Micos company\ftMicos by modifying commercial products; they are equipped with stepping motors ``Vexta NanoStep RFK Series 5-Phase Microstepping System'' produced by the Oriental Motor company\ftOriental. The motors are driven by a 4-axis ``FlexMotion PCI-7344'' board provided by National Instruments\ftNational and inserted into the host PC.
   
The ESS uses a Micos ``MS-8'' scanning table with 20.5 cm range in both directions. The coordinates are read out by two linear encoders with a resolution of $0.1 \,\mathrm{\mu m}$. External optical limit switches are mounted on each axis and manually set.
   
\begin{figure*}[ht]
 \begin{center}
  \resizebox{1.\textwidth}{!} {
    \includegraphics[height=4cm,clip]{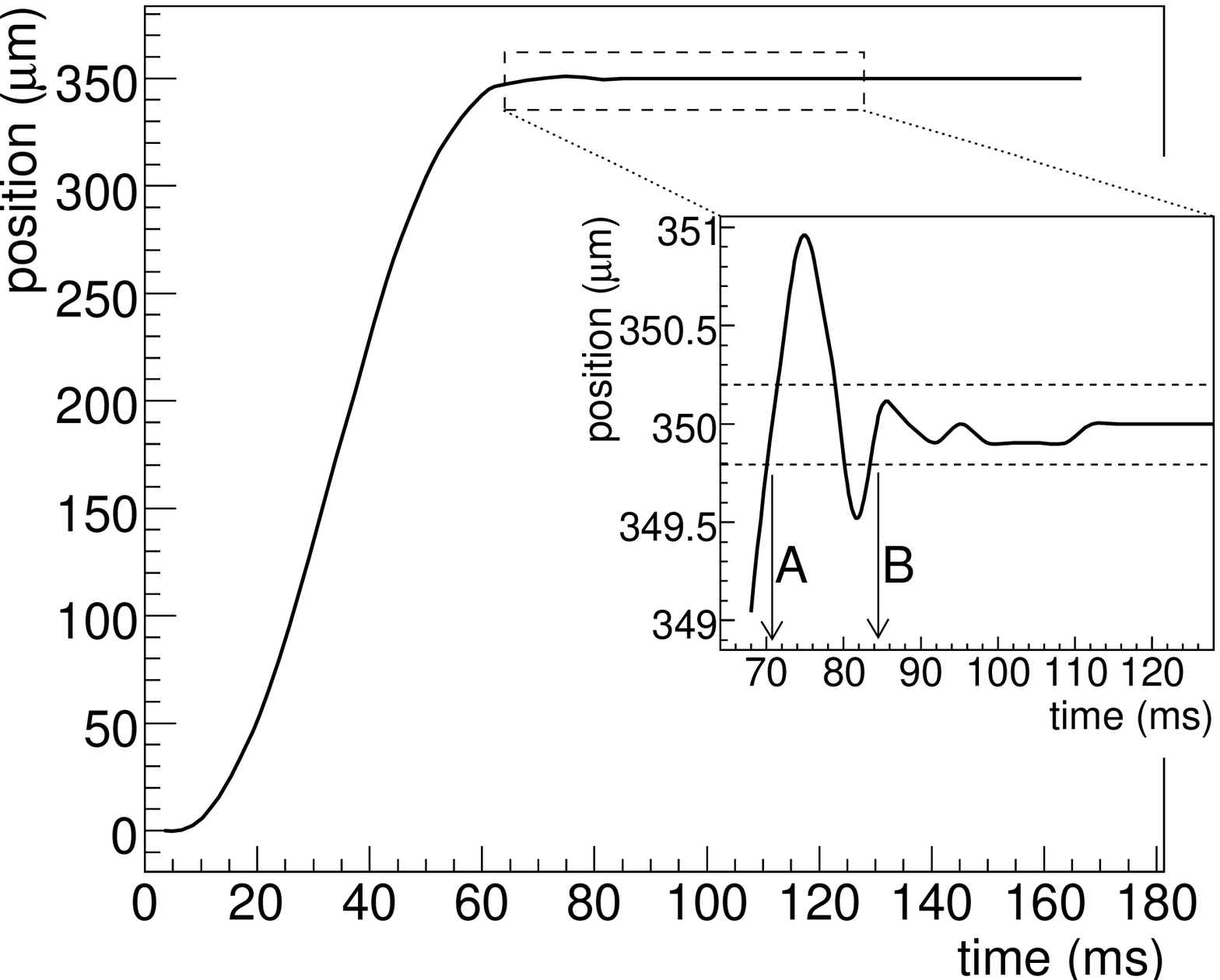}
    \includegraphics[height=4cm,clip]{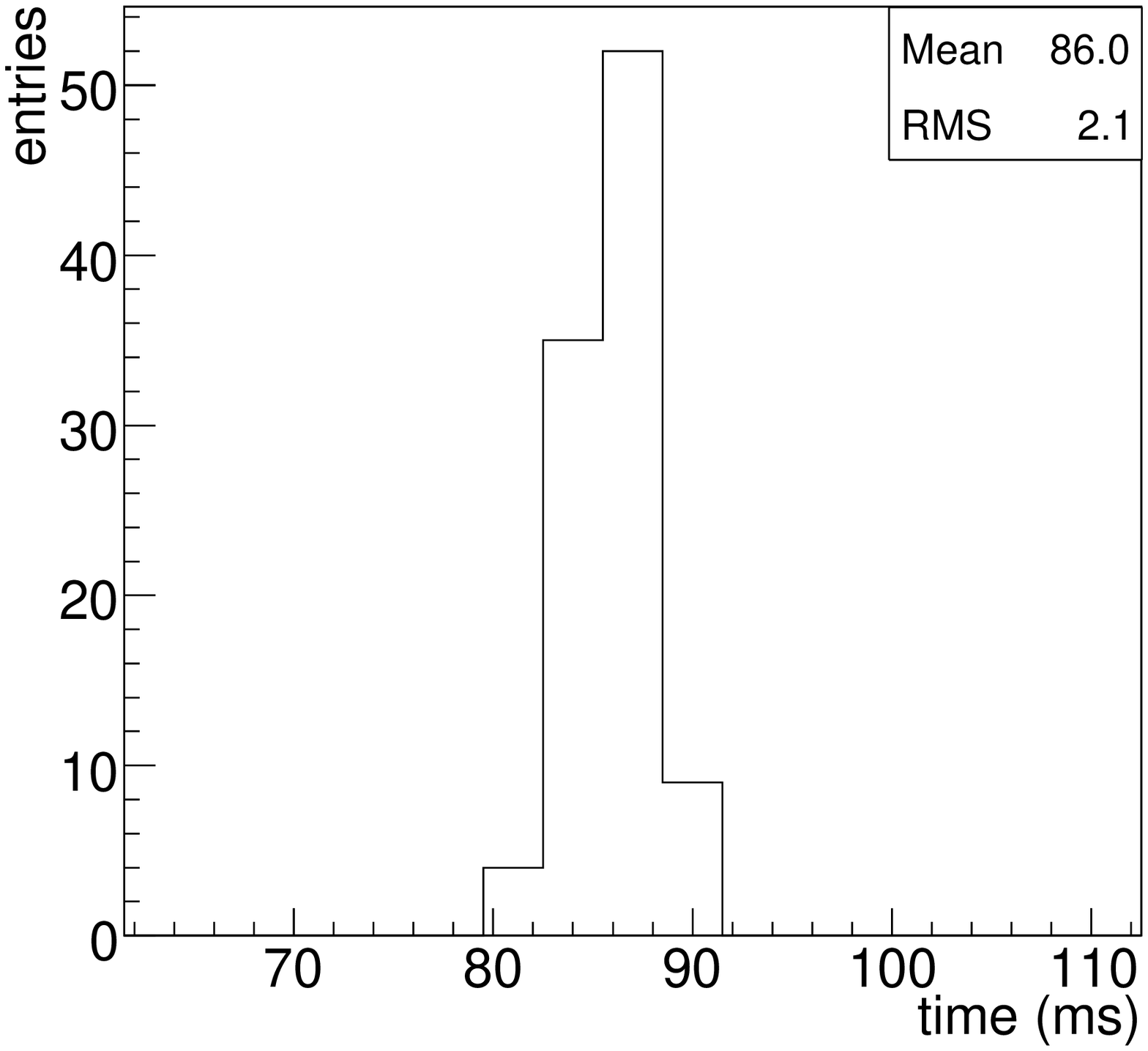} }
    (a) \hspace{0.5\textwidth} (b)
  \caption{(a) The time evolution of the stage position during the field of view change. (b) Measurement of the total time needed for a horizontal displacement which corresponds to the time interval from 0 to the arrow B in the left panel (average total time of 86 ms). \label{fi:Mov_time} }
 \end{center} \spaceafterfigurecaption
\end{figure*}

The motion of the horizontal stage (maximum speed, acceleration, deceleration, ...) was set in order to minimize the time needed to move from one field of view to the next (typically $\sim350 \,\mathrm{\mu m}$). The total displacement time is given by the sum of the \textit{rise time} and the \textit{settling time}. The former is the time to first reach the ``target point'', the latter is the time needed to wait for the oscillations to be damped to a pre-defined acceptable level which, for our working conditions, is $\pm0.2$\,$\mu$m, a value smaller than one image pixel (0.3\,$\mu$m). The  measured time profile for the X axis, for a displacement of $350\,\mathrm{\mu m}$, is shown in Fig.~\ref{fi:Mov_time}a, and the corresponding total displacement time distribution is shown in Fig.~\ref{fi:Mov_time}b. The X displacement can be safely considered concluded within $\sim100$ ms (average value: $70+16=86$ ms). The time needed for the Y axis displacements is larger ($\sim140$ ms) due to the scanning table design: the Y movements involve the whole table, while the X movements involve only a lighter part of the table. Therefore, the scanning procedure minimizes the number of Y displacements.

\begin{figure}[th] 
  \begin{center}
    \resizebox{0.5\textwidth}{!} {
      \includegraphics[clip]{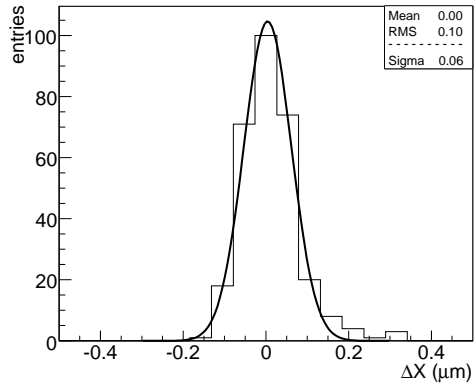}  }
    \caption{Repeatability to achieve a commanded position over many displacements in the X direction. \label{fi:grainaccu} }
  \end{center} \spaceafterfigurecaption
\end{figure}

The repeatability to reach a commanded position was evaluated by moving the stage in the X (or Y) direction, coming back each time to the starting position. This test was carried out with the motion parameters indicated by the settling time optimization and a step corresponding to a field-of-view change ($\sim350$~$\mu$m). The grey level ``center of mass'' of an emulsion grain was used as reference position; the results are shown in Fig.~\ref{fi:grainaccu}. 
The RMS of the distribution is $< 0.1 \, \mathrm{\mu m}$.  The data of Fig.~\ref{fi:grainaccu}, as well as those of Figs.~\ref{fi:align}b and \ref{Fig:ResAng}, were fitted to a gaussian form (solid lines); the fits, excluding some tails, are generally good and their mean values and sigmas are essentially equal to the experimental mean values and RMSs.

\subsection{Vertical stage}
The vertical stage used by the ESS is the Micos ``LS-110'' model. It is equipped with a linear encoder (resolution $0.05 \,\mathrm{\mu m}$) and limit switches. During data taking, the vertical stage moves at constant speed calculated by taking into account the camera frame rate, the number of desired frames and the emulsion thickness ($\ThickEmu \mathrm{\mu m}$). With a frame rate of about 400 frames/s and 15 levels per emulsion layer, each image is acquired at a vertical distance of about $3\,\mathrm{\mu m}$; the resulting speed is about $1150 \;\mathrm{ \mu m/s}$; the time needed to scan an emulsion layer is about 55 ms (including the time for acceleration, deceleration and synchronization with the host).
   
\begin{figure*}[tb]
  \begin{center}
     \resizebox{0.9\textwidth}{!} {
       \includegraphics[clip]{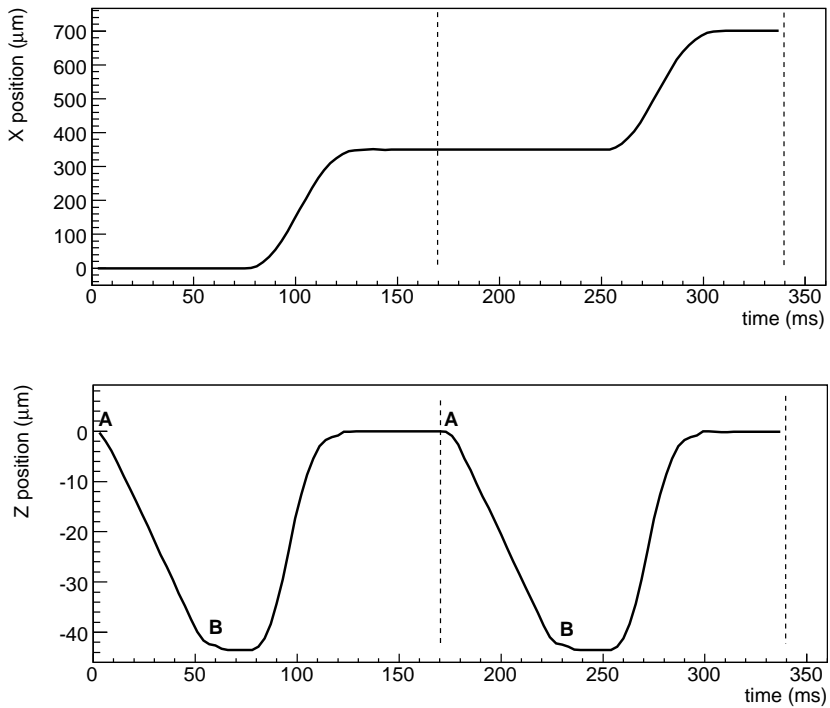}}
     \caption{The ESS X and  Z movements  during 2 data taking cycles. The vertical scans (grabbing of 15 frames at different Z levels) are performed from A to B ($\sim 55$ ms). The horizontal displacement in X lasts $\sim 90$ ms; adding few ms before and after frame grabbing leads to a full cycle of $\sim170$ ms. \label{fi:Scanning_cycle} }
  \end{center} \spaceafterfigurecaption
\end{figure*}

The time for a cycle is thus obtained by adding the time for horizontal displacement (it includes the time the vertical stage takes to reach its starting position) and the time needed for the data acquisition in Z. The insertion of a synchronization time of a few milliseconds before and after the frame grabbing brings to a $\sim170$ ms cycle time (Fig.~\ref{fi:Scanning_cycle}). As explained in Sect. \ref{Sec:acq}, this value is adequate to reach the requested scanning speed.

%______________________________________________________________________________
%
\section{Optical system}

\begin{figure*}[htb]
  \begin{center}
    \resizebox{0.65\textwidth}{!} {
      \includegraphics[clip]{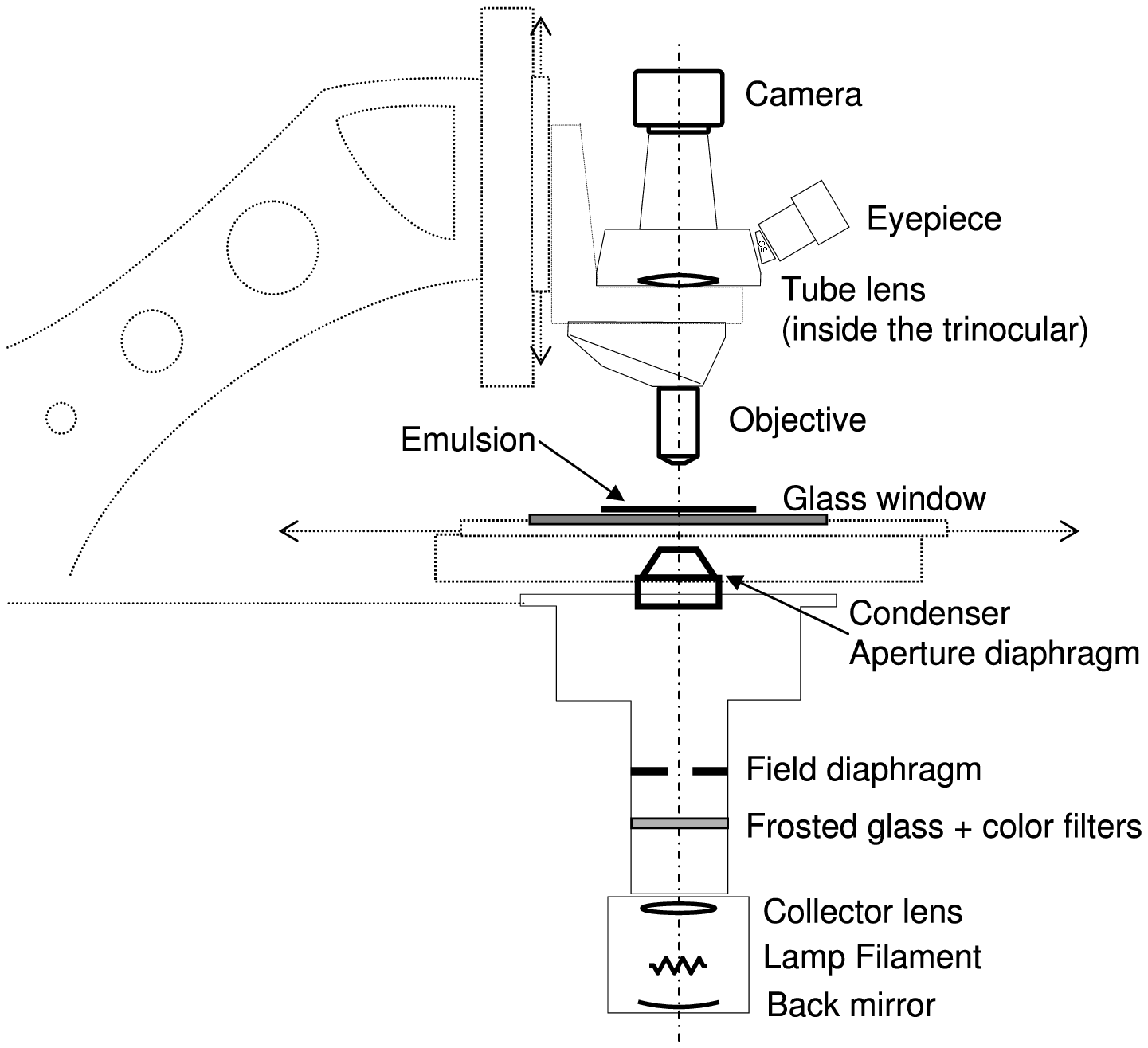}  }
    \caption{Schematic layout of the ESS microscope optical system. \label{fi:optical_system} }
  \end{center} \spaceafterfigurecaption
\end{figure*}

A schematic layout of the ESS optical system is shown in Fig.~\ref{fi:optical_system}.
   
The features and the quality of the objective have to match the requests of the scanning. The alignment of stages, light source, glass window, diaphragms and lenses must be within 1 mrad in order to satisfy the resolution requested by the OPERA experiment (2 mrad) \cite{OPERA1}. The intensity and color of the light have to be adjusted to maximize the image contrast. In the following subsections the choice of the different elements are discussed.

\subsection{Objective}
The performances of the objective should cope with the requirements of a sub-micron resolution, the need to focus at different Z depths, a magnification of few pixels per micron. An objective is characterized by the numerical aperture ($N.A.$), the working distance ($W.D.$) and the magnification ($M$). Moreover, an objective is designed to operate (or not) in an oil-immersion set-up.
   
The $N.A.$ defines the ultimate image resolution (the minimal distance between two points seen as separate) that can be achieved by the objective. Since sub-micron resolution is needed, the objective is required to have $N.A.> 0.8$\,\cite{Smith}.
   
Given the overall thickness of the emulsion layers and of the plastic support $(\ThickEmu+\ThickBase+\ThickEmu)\,\mathrm{\mu m}$, a $W.D.>0.3$ mm is required.
   
When the system scans the bottom emulsion layer, the whole plastic support and the top emulsion layer lay between the objective front lens and the focal plane, for a total thickness of 0.3~mm. For the scanning of the top emulsion layer there is no intermediate medium. The main effect of changing an intermediate medium thickness 
%with respect to the objective nominal value 
is to overcorrect or undercorrect the spherical aberration\,\cite{Smith}. An oil-immersion objective is the best choice since the oil, the emulsion and the plastic support have the same refractive index ($\sim1.5$) and therefore the optical path is almost homogeneous.
   
The objective magnification depends on the image sensor  size  because an image with at least a few pixels  per micron  is needed. In the case of 20 mm wide megapixel sensors  (see discussion in Section\,\ref{Camera}), an objective  with $M>$\,40  is needed. However, the magnification should not be much larger, in order not to  reduce the microscope speed.
   
These requirements are severe and only few objectives fulfilling all of them can be found on the market. Our choice was the Nikon CFI Plan Achromat 50x oil,  $N.A.=0.9$, $W.D.=0.4$~mm used in infinity-corrected system with a
tube lens housed in its trinocular tube. 
   
It is worth mentioning that for the OPERA experiment, where a very large number of emulsion films has to be scanned daily, the oil immersion objective is not the best practical choice. For this reason, an R\&D activity on dry objectives is still in progress. 

\subsection{Illumination}
A transmitted illumination system is placed below the scanning table. It was developed jointly with Nikon-Italy; it was designed to obtain the Koehler configuration\,\cite{Smith}. The light comes from a tungsten halogen lamp with a computer controlled power supply. The image of the lamp filament is focused by a lens (collector) on the aperture diaphragm of a condenser which concentrates the light into a cone that illuminates the emulsion sheet. A second diaphragm (field diaphragm) is adjusted to prevent emulsion illumination (and also heating) outside the field of view. The condenser numerical aperture should match that of the objective in order to have a wide illumination cone and an optimal optical resolution. The emulsion holder, described in the next section, requires that the condenser working distance is at least few mm. 
   
The final choice was a Nikon achromatic condenser with $N.A.=0.8$ and $W.D.=4.6$ mm. A green filter and a frosted glass diffuser are inserted into the light path to obtain an illumination as uniform as possible over the entire field of view and to maximize the optical resolution.

\subsection{The emulsion holder and the alignment}
The angular resolution needed for the ESS is few mrad; the systematic error introduced in the angular measurement by non-planarity of the glass window (which holds the emulsion) and by misalignments between the optical components and mechanical stage, has to be kept well below 1 mrad.
   
The glass window, equipped with a vacuum system to keep the emulsion steady during the scanning, is 4~mm thick (this is compatible with the condenser working distance). It has a thickness tolerance of less than 10~$\mu $m per 10~cm length and its deviation from the parallelism is smaller than 1 mrad; the flatness is of a few fringes per inch ($\sim0.5$~$\mu $m per 1~cm). A 1~mm wide groove in the glass along the emulsion edge is connected to a vacuum pump.

\begin{figure*}[htb]
  \begin{center}
  \resizebox{0.47\textwidth}{!} {
          \includegraphics[clip]{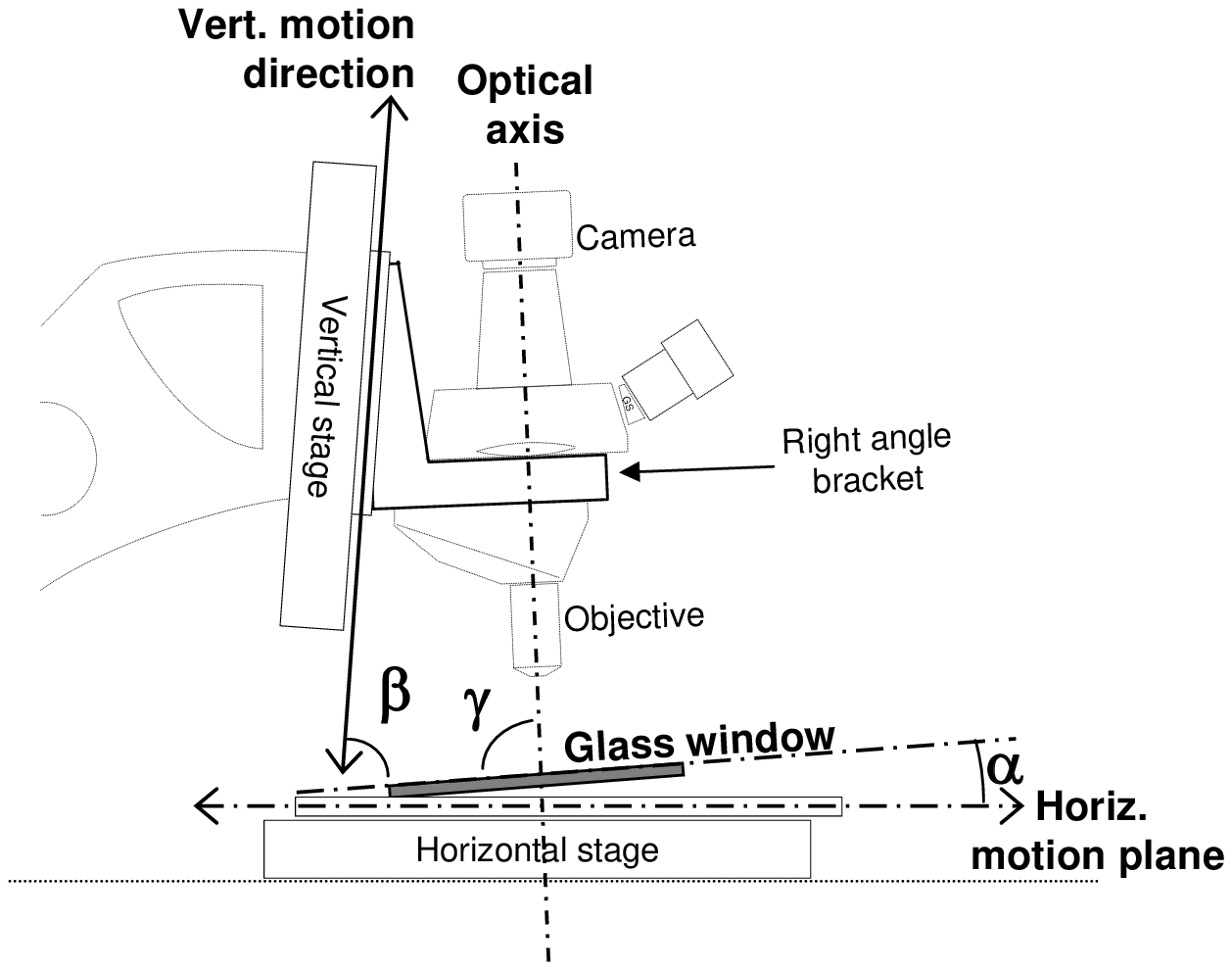}}
  \resizebox{0.51\textwidth}{!} {
        \includegraphics[clip]{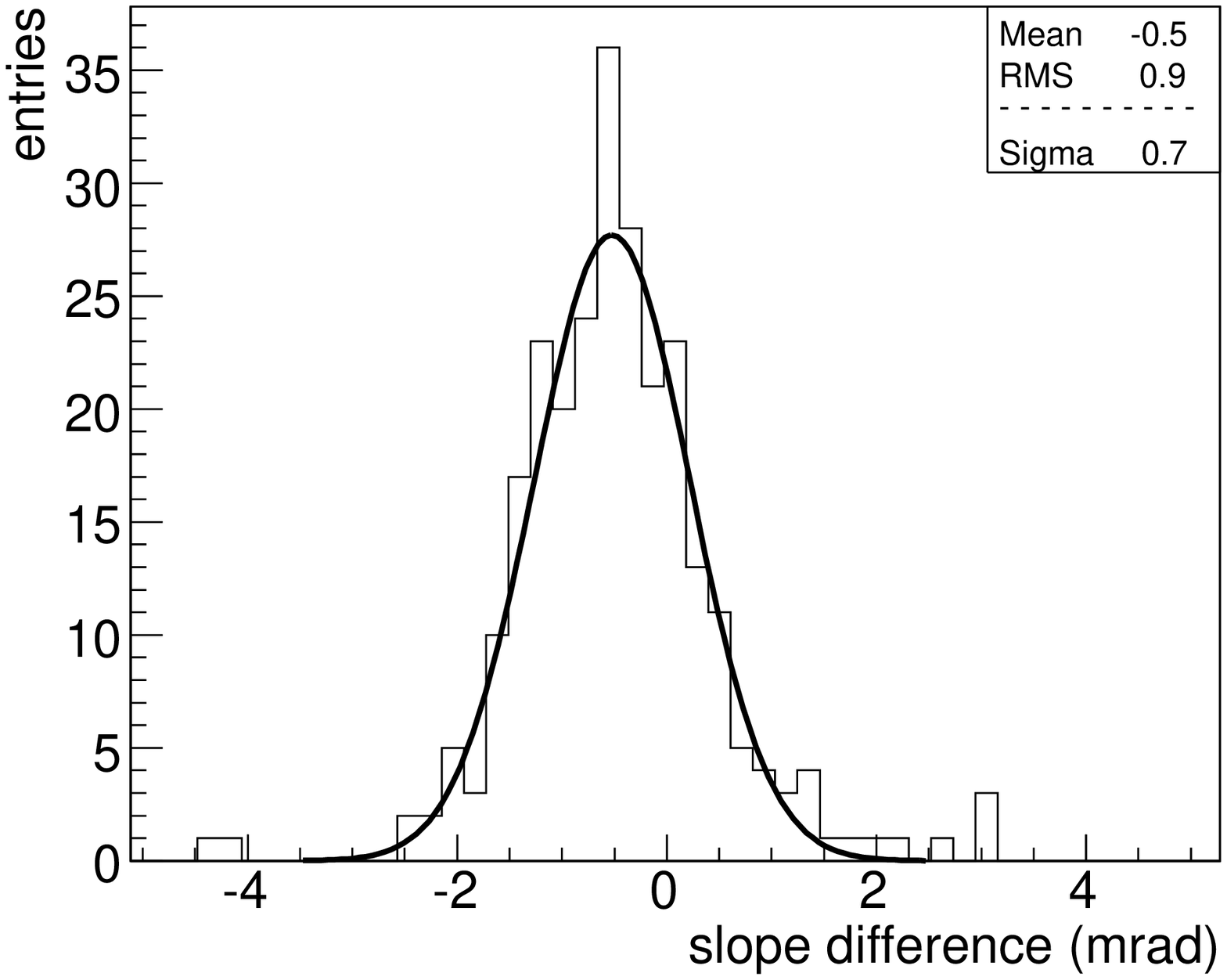}  }
           {\small(a)\hspace{0.5\textwidth}(b)}
  \caption{(a) The horizontal and vertical motion directions and the
   optical axis are aligned with reference to the glass window. The angles
   $\alpha$, $\beta$ and $\gamma$ are measured using digital comparators
   and an autocollimator. (b) The distribution of the difference between 
   measured and reference track slopes. The reference slopes have been obtained by averaging 
   the 2 slopes measured before and after a $180^\circ$ horizontal rotation of the emulsion sheet; the residual mean value of 0.5 mrad is a good estimate of the systematic uncertainty arising from possible misalignments.\label{fi:align} }
 \end{center} \spaceafterfigurecaption
\end{figure*}
   
The stages and the optical axis are aligned with respect to the glass window (used as a reference plane). The angles between the glass window and the horizontal and vertical motion directions (angles $\alpha$ and $\beta$ in Fig.~\ref{fi:align}a) are adjusted with an accuracy $\leq 0.1$~mrad using a digital micrometric comparator. The ``right angle bracket'' in Fig.~\ref{fi:align}a is aligned using an autocollimator and the final alignment of the optical axis is $\leq 0.4$~mrad (angle $\gamma$ in Fig.~\ref{fi:align}a). All the optical components shown in Fig.~\ref{fi:optical_system} are aligned using a centering telescope. 
   
In order to estimate the systematic angular uncertainty, some emulsions were vertically exposed to a 10 GeV $\pi^-$ beam. Fig.~\ref{fi:align}b shows the distribution of the difference between measured and reference track slopes. The reference slopes have been obtained by averaging the 2 slopes before and after a $180^\circ$ horizontal rotation of the emulsion sheet; the residual mean value of 0.5 mrad is a good estimate of the systematic uncertainty arising from possible misalignments.
   
%______________________________________________________________________________
%
\section{The acquisition system\label{Sec:acq}}

\subsection{Camera\label{Camera}}

The goal of 20~$\mbox{cm}^2$/h scanning speed requires a frame acquisition time $<4$ ms and megapixel resolutions.
  
The ESS is equipped with a Mikrotron MC1310\ftMikrotron\ high-speed megapixel CMOS camera with Full Camera Link interface. Its image sensor is the Micron MT9M413\ftMicron\ which delivers up to 10-bit monochrome $1280\times1024$ images at over 500 frames per second. The sensor size is 20 mm (along the diagonal) and its pixels are $12{\times} 12$ $\mu $m$^{2}$ large. This choice follows a detailed R\&D work with a custom FPGA-driven camera equipped with the same sensor. This study allowed to verify the sensor suitability in terms of speed, stability and image quality and to study and implement techniques like look-up tables, image correction, etc.
 
\subsection{Grain image\label{Sect:GrainImage}}
\begin{figure*}[tbp]
 \begin{center}
  \resizebox{1.0\textwidth}{!} {
    \includegraphics[clip]{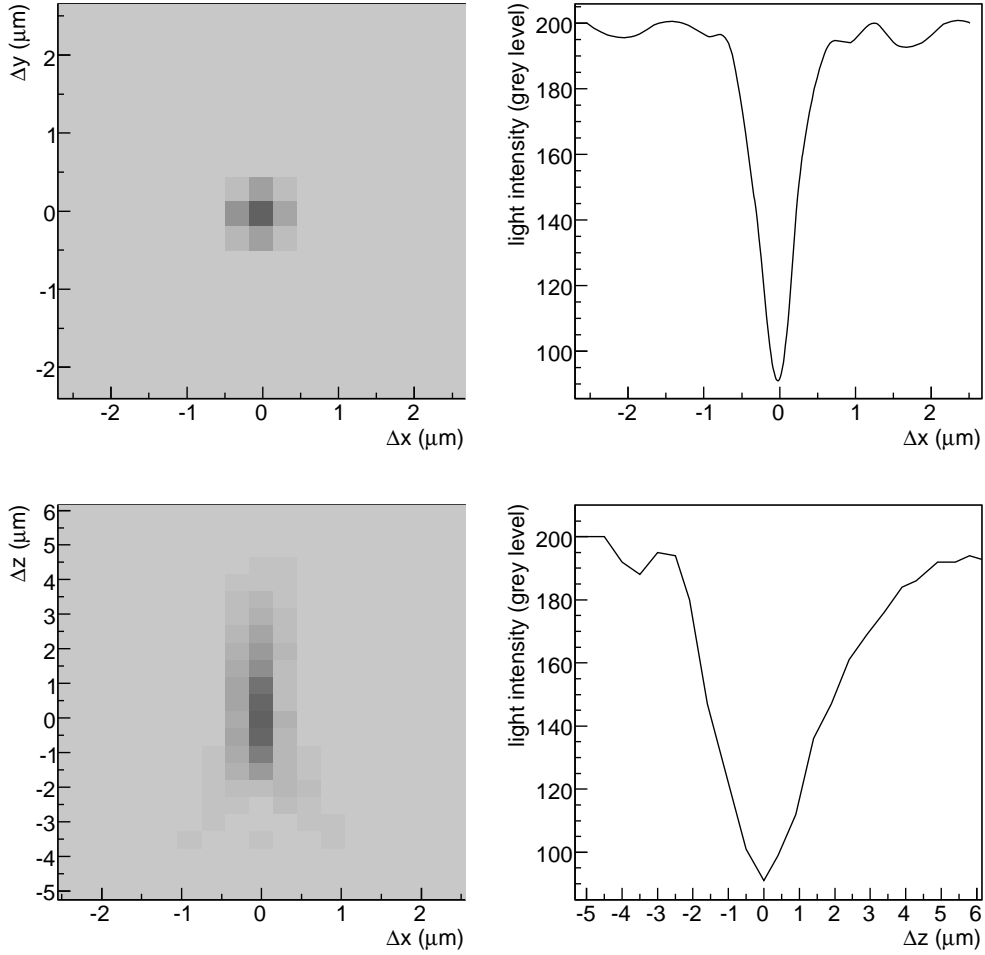} }
  \caption{Images and grey level profiles of a grain at $\Delta X =0$. On the top, a horizontal ($X-Y$) image of a focused grain and the profile along x is shown. On the bottom, the reconstruction of the same grain along the optical axis ($X-Z$ plane) and its vertical profile are shown; notice the asymmetry due to residuals of uncorrected spherical aberration. \label{fi:grain_xy} \label{fi:grain_z}}
 \end{center} \spaceafterfigurecaption 
\end{figure*}

The optical system and the CMOS camera provide a suitable grain image acquisition in terms of stability, photometric dynamics and resolution. The sensor size, the objective magnification and the setup conditions give a field of view of about $390{\times}310$ $\mu$m$^2$ and image pixels of about $0.3{\times}0.3$ $\mu$m$^2$. Consequently, the image of a focused grain is $\sim10$ pixels large. Fig.~\ref{fi:grain_z} shows a horizontal grain image (top) and the reconstructed vertical image (bottom) (obtained by taking 20 images in $\sim10$ $\mu$m at a Z distance of 0.5 $\mu$m one from the other). As indicated by the grey level profiles on the left, a focused grain can be acquired with sub-micron resolution in the horizontal ($XY$) plane while its profile is spanned over 3-4 $\mu$m along the vertical ($Z$) direction.
 
\subsection{Image acquisition and on-line processing board}
The frame grabber and the image processor are integrated in the same board, a Matrox Odyssey Xpro\ftMatrox, specifically designed to perform on-board image processing.
The on-board processor is a Motorola G4 PowerPC supported by a Matrox custom parallel processor specifically designed to quickly perform local and point-to-point operations. It is equipped with a 1 GB DDR SDRAM memory; the internal I/O bandwidth can achieve over 4 GB per second transfer rate, while the external rate reaches 1 GB per second. A Full Camera Link connection allows an acquisition rate from the camera of up to 680 MB/s.
   
At present, a camera frame rate of 377 fps and 8-bit grey level images are used corresponding to an
acquisition rate of 471 MB/s. By acquiring 15 frames per \ThickEmu $\mu$m
emulsion layer, an acquisition time of about 40 ms is needed for each
field of view. Considering a synchronization time of 15 ms, a mean
time of $\sim90$ ms for the field of view change, a field of view of
about $390{\times}310$ $\mu$m$^2$ and a superimposition between contiguous fields of 30 $\mu$m, a scanning speed  of about 22 cm$^2$/h is
obtained. 
The effective scanning speed is a bit lower ($\sim20$ cm$^2$/h) because sometimes the microscope has to scan the full sheet thickness to find the emulsion surfaces
(focusing). For precision measurements \cite{PreMea} or in particular applications, the microscope can be used with other objectives and camera resolutions; the scanning speed changes accordingly. 
   
The selected exposure time is limited by grain size ($\sim1$\,$\mu$m)  and by vertical resolution ($\sim3$\,$\mu$m). Therefore, the exposure time should be smaller than the time needed to span 1 $\mu$m. In our working condition the exposure  time is smaller than 0.5 ms (the exposure time depends on the microscope and it ranges between 0.15 and 0.5\,ms).
    
\begin{figure*}[htp]
 \begin{center}
  \resizebox{0.49\textwidth}{!} {
    \includegraphics[clip]{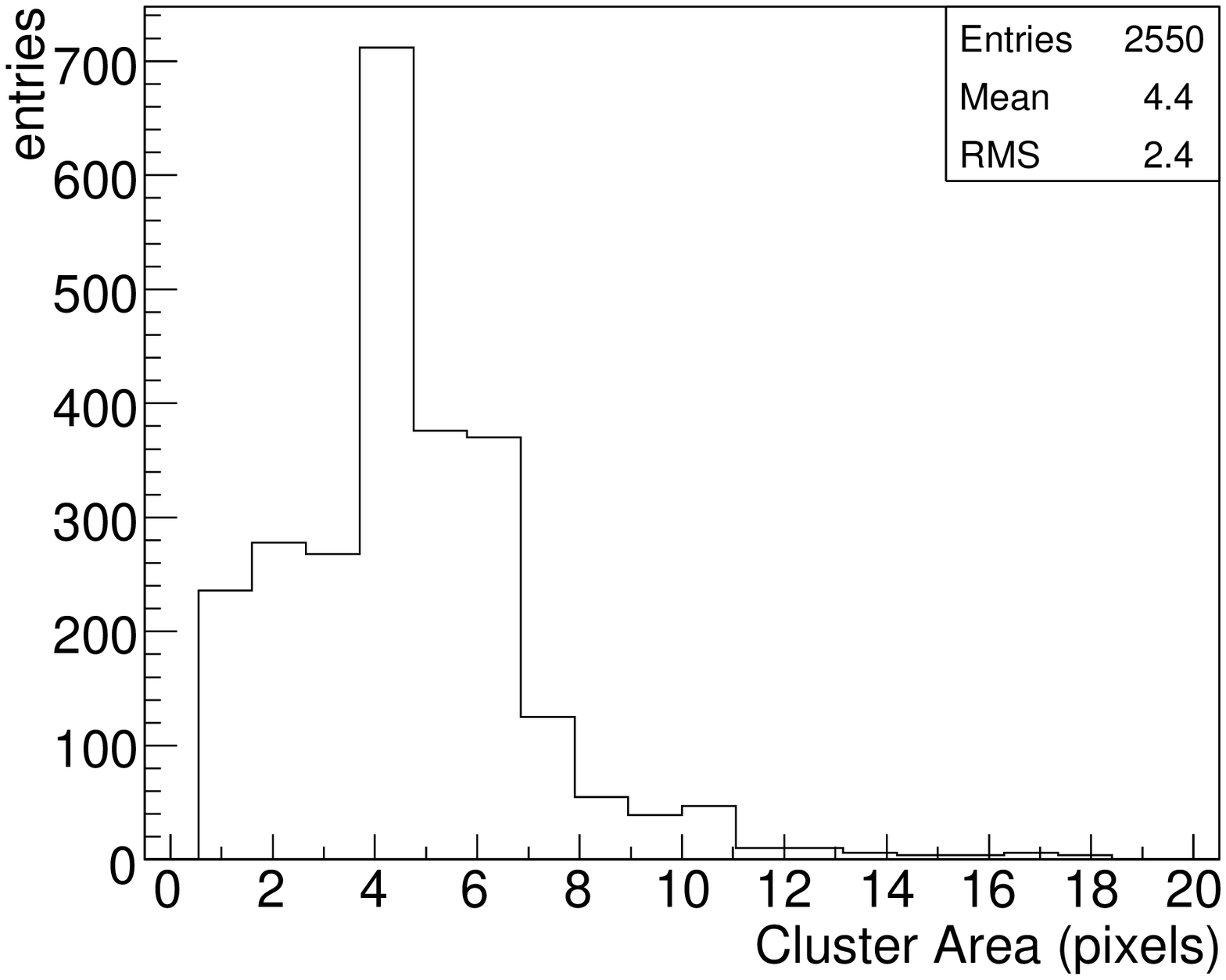} }
  \resizebox{0.49\textwidth}{!} {
    \includegraphics[clip]{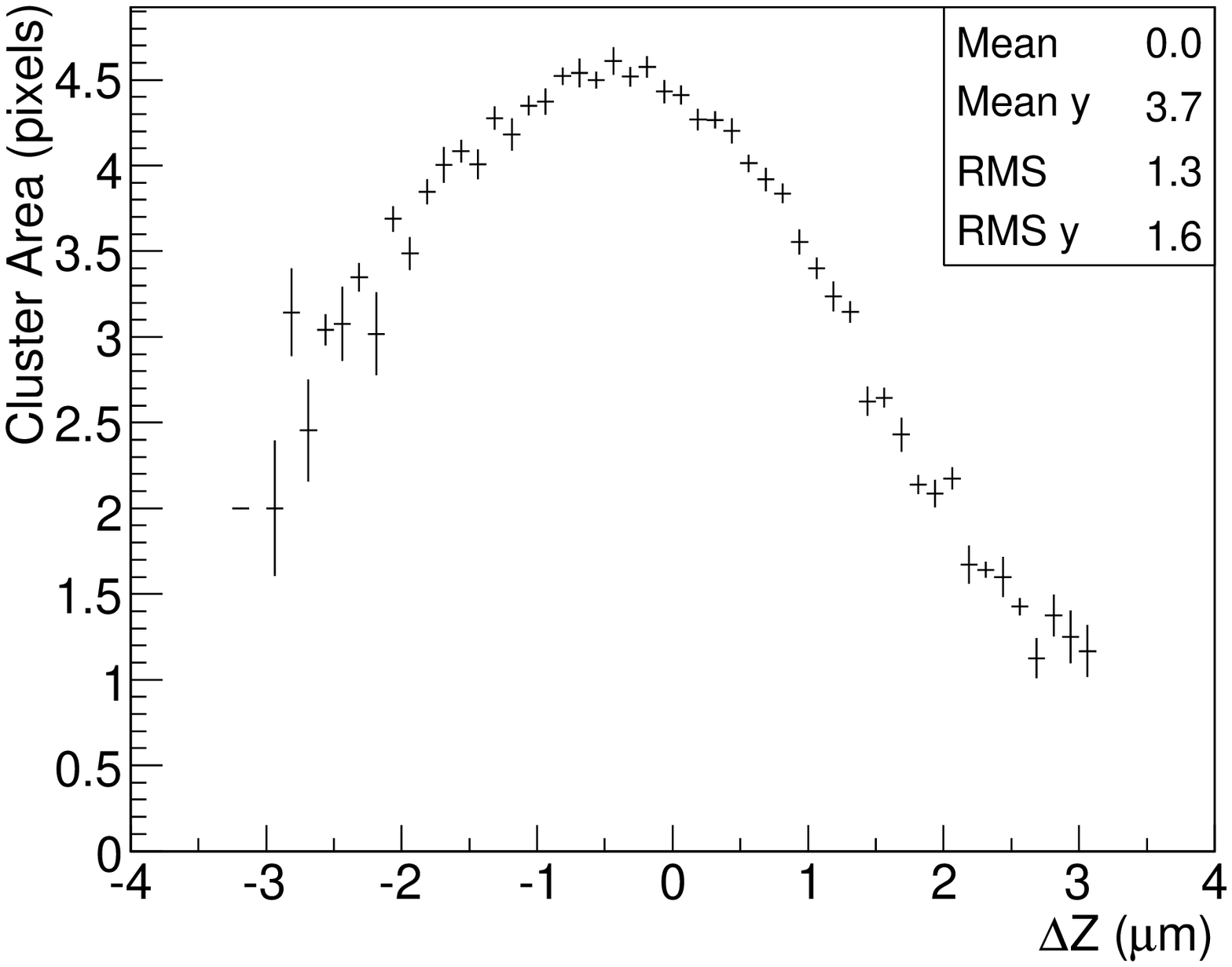} }
           {\small(a)\hspace{0.5\textwidth}(b)}
  \caption{(a) Cluster area distribution in one frame (after the image processing described in \cite{ESSsw}).\label{fi:clarea} (b) The grain image area  versus the vertical distance between the grain center-of-mass and the position where the image is grabbed; notice the asymmetry due to residuals of uncorrected spherical aberration. \label{fi:area_z}}
 \end{center} \spaceafterfigurecaption \extraspaceafterfig
\end{figure*}

Once grabbed, each image is analyzed using image processing techniques like filters, binarization and clustering (for more details see \cite{ESSsw}); the grains are recognized as clusters of black pixels. The number of clusters available for tracking depends on the threshold and on the selection cut in the cluster area. The distribution of the clusters area in one emulsion layer image is shown in Fig.~\ref{fi:clarea}a. The cluster area as a function of the distance from the grain center-of-mass is shown in the Fig.~\ref{fi:area_z}b. The efficiency and the precision in grain finding depend on the cut in the cluster area (a lower area cut yields a lower precision and a higher grain finding efficiency). Taking into account the acquisition speed and the Z level distances ($\sim 3$\,$\mu$m) the cluster area cut is chosen in order to maximize the grain finding efficiency (normally 3 or 4 pixels).

%______________________________________________________________________________

\section{Results}
\begin{figure*}[htp]
 \begin{center}
    \resizebox{0.65\textwidth}{!} {
      \includegraphics[clip]{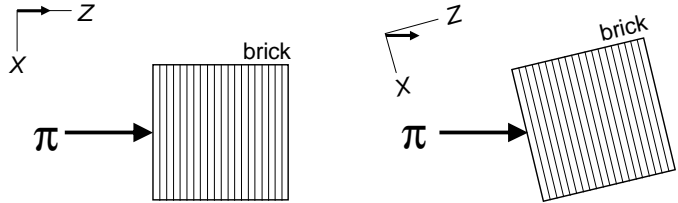}  }
  \caption{A `brick'' with 64 Fuji emulsion sheets was exposed to a 10 GeV/c $\pi^-$ beam (in the Z direction) at different
angles by rotating the brick with respect to the beam axis. \label{fi:beam}}
 \end{center} \spaceafterfigurecaption \extraspaceafterfig
\end{figure*}

A test beam exposure with emulsions was performed at CERN in June 2004 to measure the angular resolution and the efficiency.
A sample of 64 Fuji emulsion sheets was assembled manually to form a ``brick'', without lead in order to
minimize multiple scattering. 
The brick was exposed to a 10 GeV/c $\pi^-$ beam, at several different angles by rotating the brick
as shown in Fig.~\ref{fi:beam}.

After exposure, each sheet was scanned with the ESSs searching for straight sequences of grains at different Z levels with a fast tracking algorithm \cite{ESSsw}. 
The distance between two consecutive levels was chosen as $\ThickEmu\!\textrm{ }\mu\textrm{m}/15\textrm{ levels}\simeq3\textrm{ }\mu\textrm{m}$ as explained in the previous section. 
%Starting from $\sim2000$ clusters on each level and 15 levels for each 
%emulsion layer, $100\div400$ track segments (\textit{microtracks}) are found 
%in a volume corresponding to one field of view and the \ThickEmu$ \mu$m 
%thickness of the emulsion layer. 
In each field of view $200\div300$ track segments (\textit{microtracks}) are found.

\begin{figure*}[htp]
 \begin{center}
  \resizebox{0.49\textwidth}{!} {
    \includegraphics[clip]{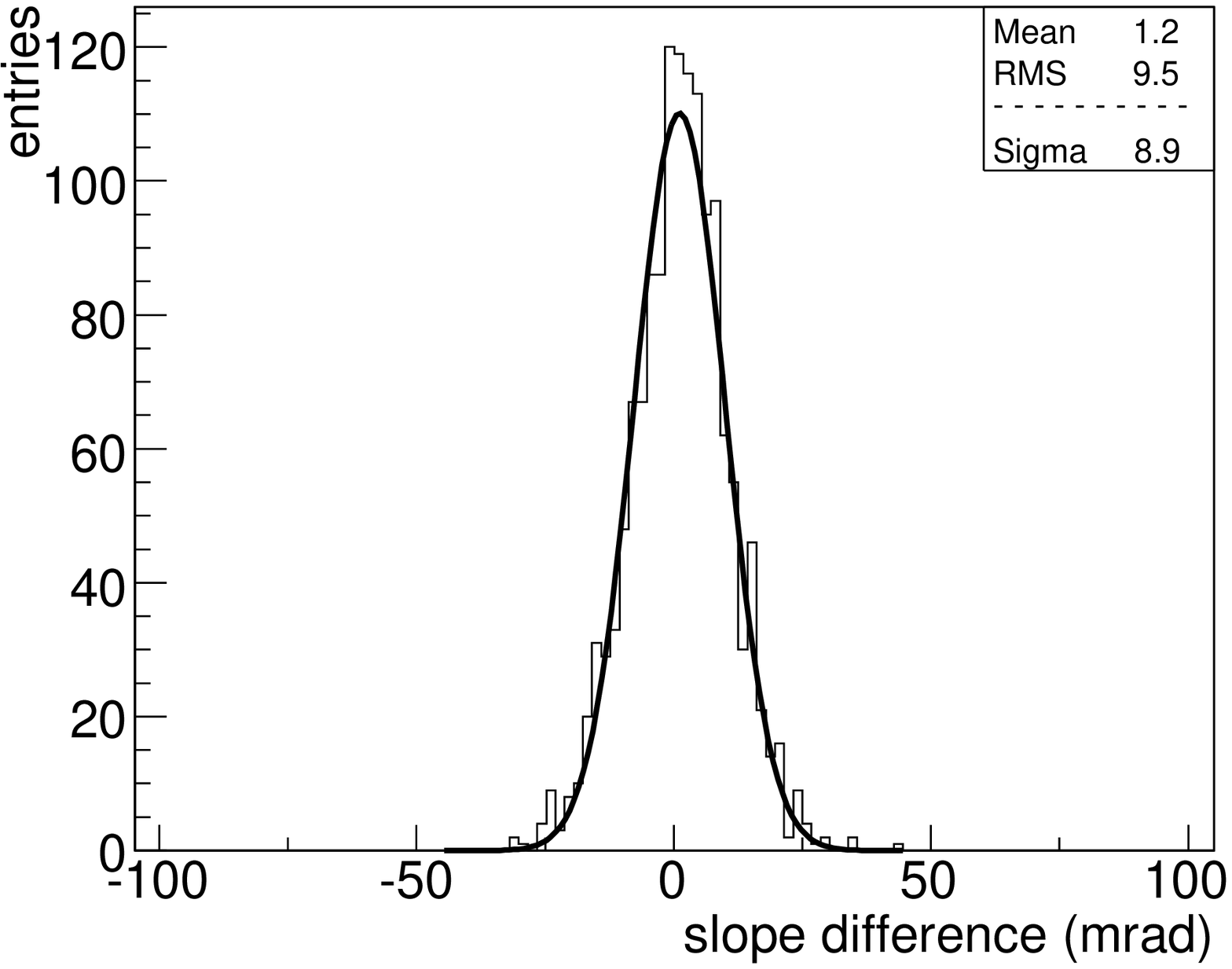} }
  \resizebox{0.49\textwidth}{!} {
    \includegraphics[clip]{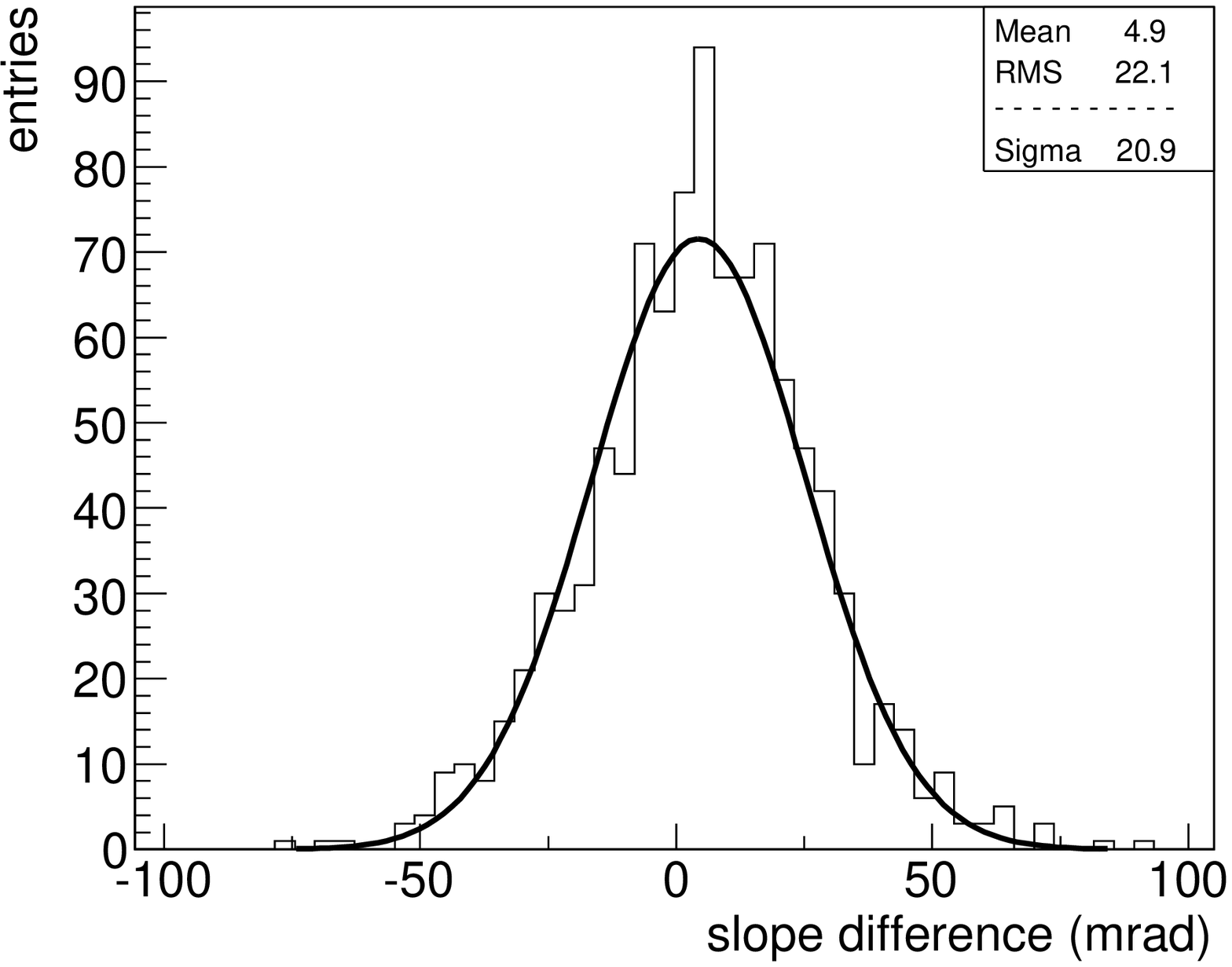} }
           {\small(a)\hspace{0.5\textwidth}(b)}
  \caption{
Residuals between microtrack slopes %(measured in the top side) 
and base track slopes from a  10 GeV/$c$ $\pi^-$ beam (a) incident vertically and (b) incident at 400 mrad.\label{Fig:ResAng}}
 \end{center} \spaceafterfigurecaption
\end{figure*}

The number of fake tracks reconstructed (mainly due to combinatorial background) is strongly reduced by connecting two microtracks across the plastic base (\textit{base tracks}) as shown in Fig. \ref{fi:Linking}b. Fig. \ref{Fig:ResAng} shows the residuals between microtrack slopes and base track slopes.

Conventional statistical selection criteria finally reduce the instrumental background at the level of 1 fake base track/cm$^2$ as measured in a not-exposed emulsion sheet.

The base track efficiency is evaluated after emulsion sheets alignment and track reconstruction in the entire brick (to recognize the \textit{volume tracks}). The  efficiency is defined as the number of measured passing-through volume tracks measured which hit one sheet with respect to the total number of passing-through volume tracks; the efficiency is greater than 90\% and corresponds to a microtrack finding efficiency of about 95\%. 

\begin{figure*}[htp]
 \begin{center}
  \resizebox{0.49\textwidth}{!} {
    \includegraphics[clip]{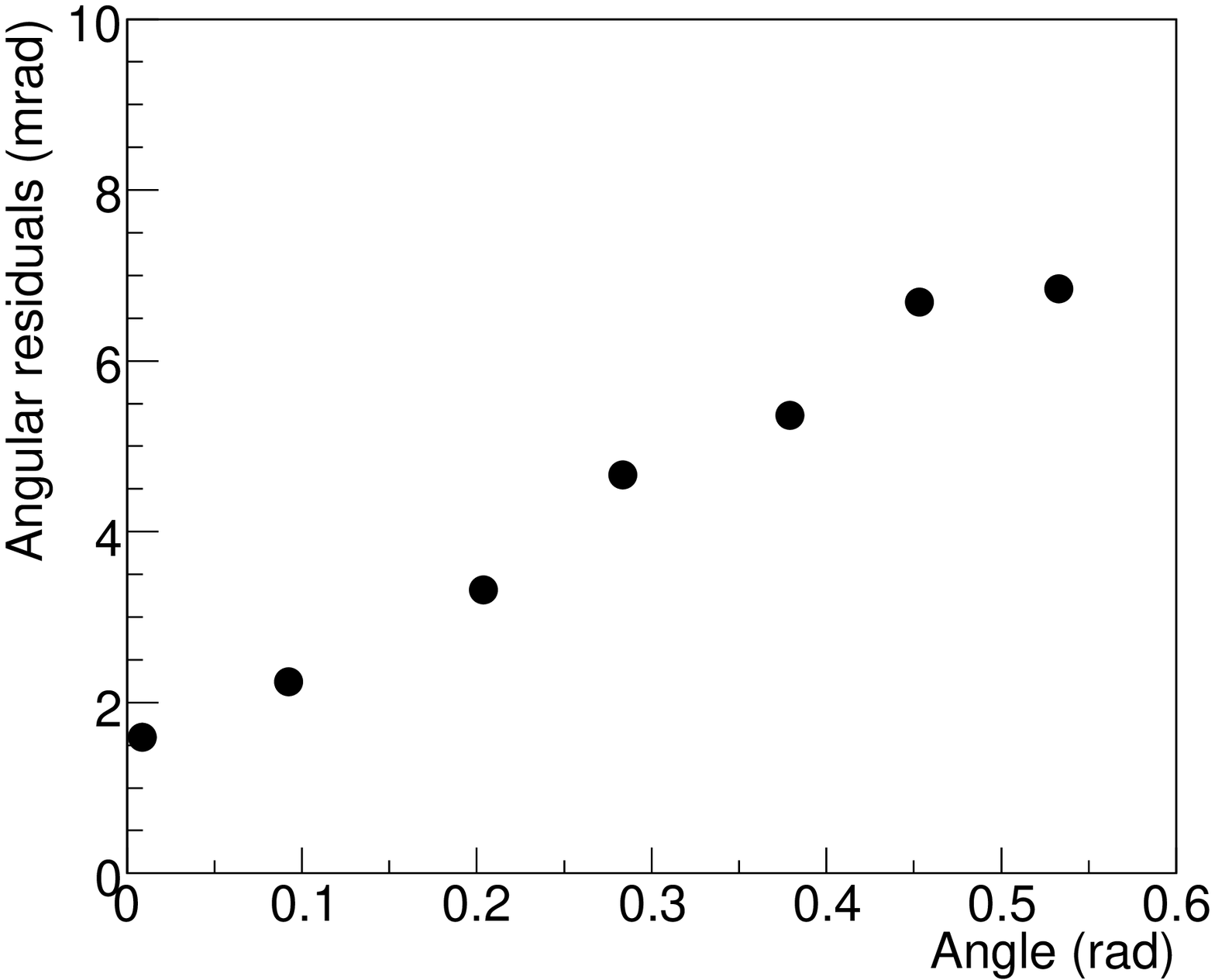} }
  \resizebox{0.49\textwidth}{!} {
    \includegraphics[clip]{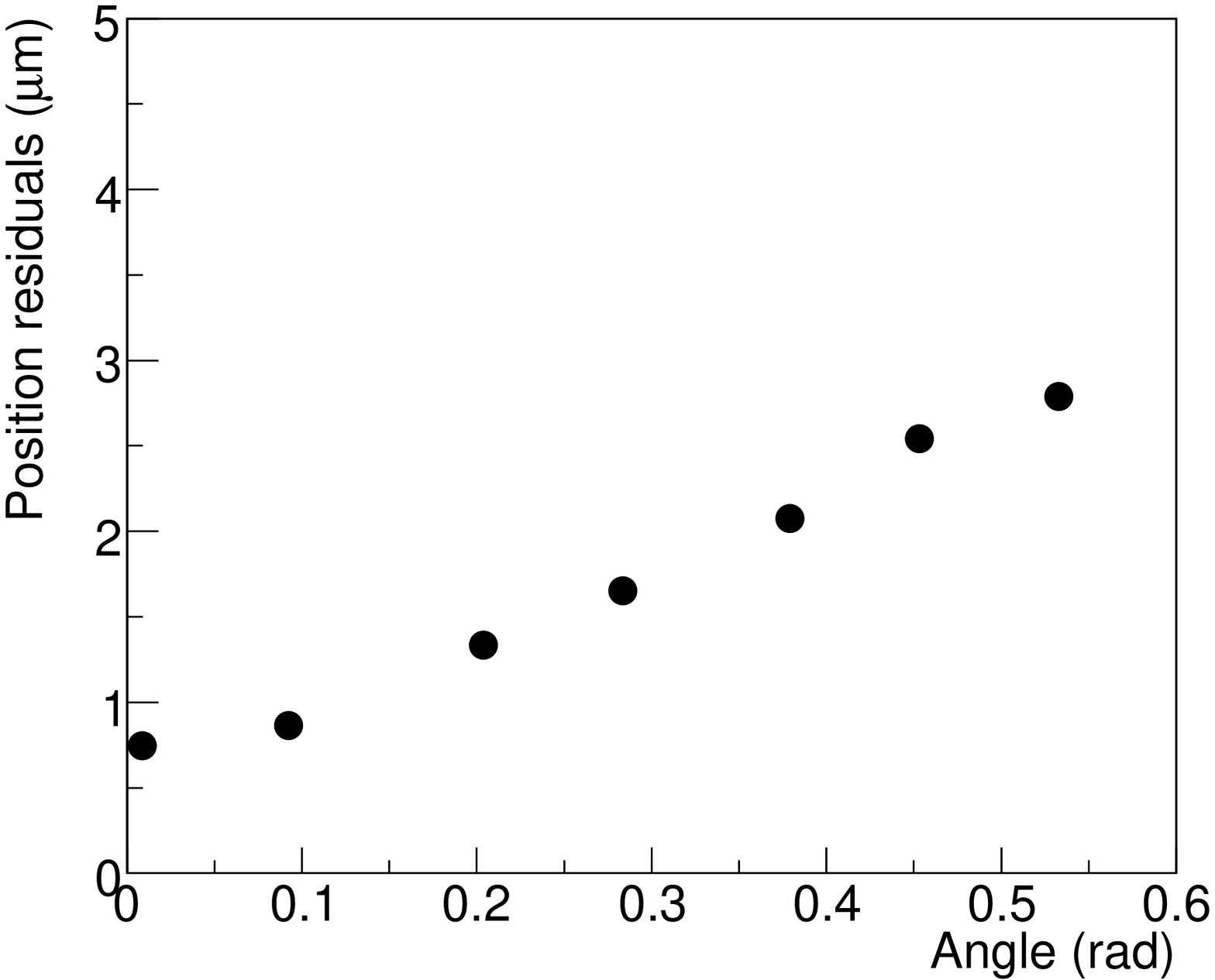} }
           {\small(a)\hspace{0.5\textwidth}(b)}
  \caption{
(a) The angular resolution of base tracks as function of the reconstructed angle. It is evaluated by comparing base track angles with respect to the volume track angles. (b) The position resolution of base tracks. The errors (that are inside the dimensions of each black point) are only statistical.\label{Fig:VolumeResAng}}
 \end{center} \spaceafterfigurecaption
\end{figure*}

The measured base track angular and position resolutions are shown in Fig. \ref{Fig:VolumeResAng}. The residuals depend on the volume track direction and range from 1.6 to 7 mrad and 0.9 to 2.5 $\mu$m in the [0,600] mrad angular range. 

The angular resolution of a base track is related to the accuracy in the measurement of the microtrack intercept positions in the base ($\Delta X\textrm{,}\Delta Y\sim0.3$ $\mu$m and $\Delta Z\sim3$ $\mu$m as described in Sect. \ref{Sect:GrainImage}); the angular resolution is one order of magnitude better than the microtrack resolution (which are spread by emulsion distortion and shrinkage effects). 

The scanning speed reached for these measurements is $\sim20$ cm$^2$/h per \ThickEmu $\mu$m thickness of emulsion layer.

%______________________________________________________________________________

\section{Conclusions}
   
The features and performances of the European Scanning System (ESS) hardware have been described. The choice of mechanical, optical and acquisition components was discussed in detail. The resulting microtrack efficiencies have been evaluated to be above 95\% in the [0, 600] mrad angular range.% with an accuracy of $\sim 1$ mrad for vertical tracks.

The ESS has reached the planned speed of $\sim20$ cm$^2$/h in an emulsion volume of \ThickEmu $\mu$m thickness. This represents an improvement of more than an order of magnitude with respect to the systems developed in the past. The scanning performances satisfy the requirements of the OPERA experiment.

About 20 ESSs have been installed in European laboratories collaborating in the OPERA experiment. Five more have been installed at the Gran Sasso Laboratory (LNGS) for the analysis of the Changeable Sheets.

%______________________________________________________________________________
%
\section*{Acknowledgements}
We acknowledge the cooperation of the members of the OPERA Collaboration and we thank many colleagues for discussions and suggestions. We gratefully acknowledge the invaluable support of the technical staff in our laboratories; in particular we thank L. Degli Esposti, M. Di Marino, V. Di Pinto, F. Fiorello, P. Pecchi, A. Ruggieri and V. Togo and for their contributions. We thank INFN for providing fellowships and grants (FAI) for non Italian citizens.

% \appendix

% \section{}
% \label{}

%______________________________________________________________________________
%

\end{document}